\documentstyle[11pt]{article}
\begin{document}

\title{\textbf{The Effect of Twisted Magnetic Field on the Period Ratio $P_1/P_2$ of Nonaxisymmetric MHD Waves}}
\author{K. Karami$^{1,2}$\thanks{KKarami@uok.ac.ir} , K. Bahari${^3}\thanks{K.Bahari@razi.ac.ir}$\\$^{1}$\small{Department of Physics,
University of Kurdistan, Pasdaran Street, Sanandaj,
Iran}\\$^{2}$\small{Research Institute for Astronomy $\&$
Astrophysics of Maragha (RIAAM), Maragha,
Iran}\\$^{3}$\small{Physics Department, Faculty of Science, Razi
University, Kermanshah, Iran}}

\maketitle

\begin{abstract}
The nonaxisymmetric magnetohydrodynamic (MHD) modes in a zero-beta
cylindrical compressible thin magnetic flux tube modelled as a
twisted core surrounded by a magnetically twisted annulus, both
embedded in a straight ambient external field is considered. The
dispersion relation is derived and solved analytically and
numerically to obtain the frequencies of the nonaxisymmetric MHD
waves. The main result is that the twisted magnetic annulus does
affect the period ratio $P_1/P_2$ of the kink modes. For the kink
modes, the magnetic twist in the annulus region can achieve
deviations from $P_1/P_2 = 2$ of the same order of magnitude as in
the observations. Furthermore, the effect of the internal twist on
the fluting modes is investigated.
\end{abstract}

\noindent{\textit{Subject headings:} MHD --- Sun: corona --- Sun:
magnetic fields --- Sun: oscillations}

%\noindent{{\bf Key words:}~~~Sun: corona -- Sun: magnetic fields --
%Sun: oscillations}
%-----------------------------------------------------------------------------------------------
\clearpage
\section{Introduction}
Transverse coronal loop oscillations triggered by explosive events,
such as flares or filament eruptions, were first identified by
Aschwanden et al. (1999) and Nakariakov et al. (1999) using the
observations of TRACE (Transition Region And Coronal Explorer).
These oscillations have been interpreted as the kink MHD modes of a
cylindrical coronal flux tube by Nakariakov et al. (1999).

One of the important tools in the coronal seismology is
determination of the period ratio $P_1/P_2$ between the period $P_1$
of the fundamental mode and the period $P_2$ of its first harmonic.
The deviation of the period ratio from its canonical
harmonic value of 2 has been observed in coronal loop oscillations.
Verwichte et al. (2004), using the observations of TRACE, have
identified the fundamental and its first harmonic of the transverse
kink mode in two coronal loops. The period ratios observed by
Verwichte et al. (2004) are $1.81 \pm 0.25$ and $1.64 \pm 0.23$.
However, these values were corrected with the improvement of the
observational error bars to $1.82 \pm 0.08$ and $1.58 \pm 0.06$,
respectively, by Van Doorsselaere, Nakariakov \& Verwichte (2007).
Also Verth, Erd\'{e}lyi \& Jess (2008) added some further
corrections by considering the effects of loop expansion and
estimated a period ratio of 1.54. All these values clearly
are lower than 2. This may be caused by different factors such as
the effects of density stratification (see e.g. Andries et
al. 2005; Erd\'{e}lyi \& Verth 2007; Karami \& Asvar 2007; Safari,
Nasiri \& Sobouti 2007; Karami, Nasiri \& Amiri 2009) and
magnetic twist (see Erd\'{e}lyi \& Carter 2006; Erd\'{e}lyi \&
Fedun 2006; Karami \& Barin 2009; Karami \& Bahari 2010) in
the loops. Note that in some cases the  period ratio is
shifted to higher values than 2. For instance, in Table 1 of Andries
et al. (2009) there are two observational examples with $P_1/P_2>2$.
This may be caused by the effect of magnetic field expansion (see
e.g. Verth \& Erd\'{e}lyi 2008; Ruderman, Verth \& Erd\'{e}lyi 2008;
Verth, Erd\'{e}lyi \& Jess 2008; Karami \& Bahari 2011). Also there
are some observational cases in which the period ratios do not show
any significant departures from their canonical harmonic values. For
instance, in Table 1 of Andries et al. (2009) there is an example
with $P_1/P_2 =2$. Also Van Doorsselaere et al. (2009) found
$P_1/P_2\approx2$ and $P_1/P_3\approx 3$ in a highly twisted loop
structure which is certainly not homogeneous and the loop structure
could be classified as a sigmoid.

The twisted magnetic tubes have been investigated in ample detail by
Bennett, Roberts $\&$ Narain (1999) and Carter \& Erd\'{e}lyi (2007,
2008). For a good review see Karami \& Barin (2009).

Ruderman (2007) studied the nonaxisymmetric oscillations of a
compressible zero-beta thin twisted magnetic tube surrounded with
the straight and homogeneous magnetic field taking the density
stratification into account. Using the asymptotic analysis he showed
that the eigenmodes and the eigenfrequencies of the kink
and the fluting oscillations are described by a classical
Sturm-Liouville problem. The main result of Ruderman (2007), which
also has been already obtained by Goossens, Hollweg $\&$ Sakurai
(1992), was that the twist does not affect the kink mode.
Note that in these analytical works, and in the present
work, the azimuthal component of the equilibrium magnetic field
$B_\phi$ is taken to be proportional to $r$. This is essentially
because it makes the governing equations easier to solve
analytically. However, this makes the equilibrium magnetic field a
very particular and artificial case study and may be far from
reality. It may be that more general $B_\phi$ equilibria do affect
the kink mode.

Karami \& Barin (2009) investigated both the oscillations and
damping of MHD surface and hybrid waves in coronal loops in the
presence of twisted magnetic field. They considered a straight
cylindrical incompressible flux tube with magnetic twist just in the
annulus and straight magnetic field in both the internal and
external regions. They showed that the frequencies and the damping
rates of both the kink and fluting modes increase when the twist
parameter increases. They obtained that the period ratio $P_1/P_2$
of the fundamental and first overtone for both the kink and fluting
surface modes are lower than 2 (for untwisted loop) in the presence
of twisted magnetic field.

Karami \& Bahari (2010) examined the effect of twisted magnetic
field on the resonant absorption of MHD waves in coronal loops. They
concluded that with increasing the twist, the ratio of the
oscillation frequency to the damping rate of the kink modes changes
from 39.3 to 43.5, which approximately is one order of magnitude
greater than the ratio reported by Nakariakov et al. (1999),
Verwichte et al. (2004), and Wang and Solanki (2004) deduced from
the TRACE data. Note that the twisted cylinder model
proposed by Karami \& Bahari (2010) was found produce too weak a
damping rate to explain the observed strong kink wave damping.

In the present work, our main aim is to investigate the effect of
twisted magnetic field on the frequencies of nonaxisymmetric MHD
waves in coronal loops to justify the deviation of the period ratio
$P_1/P_2$ from 2 observed by the TRACE. This paper is organized as
follows. In Section 2 we use the asymptotic analysis obtained by
Ruderman (2007) to derive the equations of motion. In Section 3 and
its subsections we present the dispersion relation, two reductions
to known cases, and an analytical solution. In Section 4,
we give numerical results. Section 5 is devoted to conclusions.

%-----------------------------------------------------------------------------------------------
\section{Equations of motion}
The linearized MHD equations for a compressible zero-beta plasma are
\begin{eqnarray}
\frac{\partial\delta\mathbf{v}}{\partial
t}=\frac{1}{4\pi\rho}[(\nabla\times\delta\mathbf{B})\times\mathbf{B}
+(\nabla\times\mathbf{B})\times\delta\mathbf{B}],\label{mhd1}
\end{eqnarray}
\begin{eqnarray}
\frac{\partial\delta\mathbf{B}}{\partial
t}=\nabla\times(\delta\mathbf{v}\times\mathbf{B}),\label{mhd2}
\end{eqnarray}
where $\delta\bf{v}$ and $\delta\bf{B}$ are the Eulerian
perturbations in the velocity and magnetic fields; $\rho$ is the
mass density.

The simplifying assumptions are as follows.
\begin{itemize}
\item The background magnetic field is assumed to be
\begin{eqnarray}
{\mathbf{B}}= \left\{ \begin{array}{ll}
{\mathbf{B}}_{\rm i}= \Big(0,A_{\rm i}r,B_{z{\rm i}} (r)\Big),& r < a, \\
{\mathbf{B}}_0=\Big(0,A_0 r,B_{z0} (r)\Big),& a < r < R, \\
{\mathbf{B}}_{\rm e}=\Big(0,0,B_{z{\rm e}}\Big),&r>R, \\
 \end{array} \right.
\end{eqnarray}
where $A_{\rm i}$, $A_0$, $B_{z{\rm e}}$ are constant and $a$, $R$
are radii of the core and the tube, respectively. From both
the equilibrium equation, i.e.
      $\frac{dB^2}{dr}=-\frac{2B_{\phi}^2}{r}$, and the continuity condition of the magnetic pressure across the boundaries of the
      tube, i.e. $B_{\rm i}^2(a)=B_0^2(a),~B_0^2(R)=B_{\rm e}^2 (R)$,
      the $z$-component of the equilibrium magnetic field can be obtained as
\begin{eqnarray}
\begin{array}{l}
B_{z{\rm i}}^2 (r) = B_0^2  + A_{\rm i}^2 \big(a^2  - 2r^2 \big), \\
B_{z0}^2 (r) = B_0^2  + A_0^2 \big(a^2  - 2r^2 \big), \\
B_{z{\rm e}}^2  = B_0^2  + A_0^2 \big(a^2  - R^2 \big), \\
\end{array}
\end{eqnarray}
where $B_0$ is an integration constant. The above magnetic field
configuration in the absence of the annulus is the same as the
background magnetic field considered by Ruderman (2007).
\item $\rho$ is constant along the loop but different in the interior, annulus and exterior regions and denoted by
$\rho_{\rm i}$, $\rho_0$ and $\rho_{\rm e}$, respectively.
\item We
consider the flux tube to be a cylinder and therefore implement the
cylindrical coordinates, $(r,\phi,z)$.
     \item The plasma equilibrium is fixed to
be in a steady state, i.e., without flow.
\item $t$-, $\phi$- and $z$-dependence for any of the components
$\delta{\bf{v}}$ and $\delta{\bf{B}}$ is $\exp{\{{\rm i}(m\phi+k_z
z-\omega t)\}}$. Here $k_{z}=l\pi/L$ and $L$ is the tube
length. Also $l$ and $m$ are the longitudinal and azimuthal mode
numbers, respectively.
\end{itemize}

Here like Ruderman (2007) we consider $\epsilon:=\frac{Aa}{B_0}\sim
k_za\ll 1$ which is in good agreement with the observations and also
look for the low frequency eigenmodes. Following the second
order perturbation method in terms of $\epsilon$ given by Ruderman
(2007), solutions of Eqs. (\ref{mhd1})-(\ref{mhd2}) in terms of
$\delta P=\frac{\mathbf{B}\cdot\delta\mathbf{B}}{4\pi}$, the
Eulerian perturbation in the magnetic pressure, and $\xi_r=-\delta
v_{r}/{\rm i}\omega$, the Lagrangian perturbation in the radial
displacement, for the interior and annulus regions yield
\begin{eqnarray}
\delta P(r) = \frac{r}{{m^2 }}\Big(\rho \omega ^2 - \frac{{B_0^2
}}{{4\pi}}F^2 \Big)\frac{{{\rm d}(r\xi _r )}}{{{\rm d}r}}
\nonumber\\+ \Big(\frac{{B_0 AF}}{{2\pi m}}\Big)r\xi _r,\label{P}
\end{eqnarray}
\begin{eqnarray}
\frac{{\rm d}}{{{\rm d}r}}\Big(r\frac{{{\rm d}(r\xi _r )}}{{{\rm
d}r}}\Big) - m^2 \xi _r = 0,\label{dxir}
\end{eqnarray}
where $ F = k_z+m\frac{A}{B_0}$. Equations (\ref{P}) and
(\ref{dxir}) are same as Eqs. (19) and (21), respectively, in
Ruderman (2007).

In the interior and annulus regions, solutions of Eq. (\ref{dxir})
are
\begin{eqnarray}
\xi _r (r) = \left\{ \begin{array}{lll}
 \alpha r^{m - 1},&r < a, \\
 \beta r^{m - 1}  + \gamma r^{ - m - 1},& a < r < R,\\
 \end{array} \right.\label{xir}
\end{eqnarray}
and solutions for $\delta P(r)$ are obtained from substituting Eq.
(\ref{xir}) in (\ref{P}) as
\begin{eqnarray}
\delta P(r) = \Big(\rho _{\rm i} \omega ^2  - \frac{{B_0^2F_{\rm
i}^2 }}{{4\pi }}  + \frac{{B_0 A_{\rm i} F_{\rm i} }}{{2\pi
}}\Big)\frac{{\alpha r^m
 }}{m},&r < a,\label{Pi01}
\end{eqnarray}
\begin{eqnarray}
\delta P(r) = \Big(\rho _0 \omega ^2  - \frac{{B_0^2F_0^2 }}{{4\pi
}}  +
 \frac{{B_0 A_0 F_0 }}{{2\pi }}\Big)\frac{{\beta r^m }}{m} \nonumber\\-
 \Big( \rho _0 \omega ^2  - \frac{{B_0^2F_0^2 }}{{4\pi }}  -
 \frac{{B_0 A_0 F_0 }}{{2\pi }}\Big)\frac{{\gamma r^{ - m} }}{m},& a < r < R.~\label{Pi02}
\end{eqnarray}
For the exterior region, $r>R$, we obtain
\begin{eqnarray} \frac{{{\rm d}^2
\delta P}}{{{\rm d}r^2 }} + \frac{1}{r}\frac{{{\rm d}\delta
P}}{{{\rm d}r}} - \Big(k'^2 + \frac{{m^2 }}{{r^2 }}\Big)\delta P =
0,\label{dPe}
\end{eqnarray}
\begin{eqnarray}
\xi _r (r) =-\frac{4\pi}{k'^2B_0^2}\frac{{\rm d}\delta P}{{\rm
d}r},\label{xire1}
\end{eqnarray}
where
\begin{equation}
k'^2  = k_z^2  - \frac{{4\pi\rho _e \omega ^2 }}{{B_0^2 }}.
\end{equation}
Equations (\ref{dPe}) and (\ref{xire1}) are same as Eqs. (26) and
(25a), respectively, in Ruderman (2007). In the exterior region,
$r>R$, the waves should be evanescent. Solutions are
\begin{eqnarray}
\delta P(r) = \varepsilon K_m (k'r),~~~~~~k'^2>0,\label{Pe}
\end{eqnarray}
\begin{eqnarray}
\xi _r (r) = - \varepsilon \frac{{4\pi }}{{k'B_0^2 }}K_m ^\prime
(k'r),\label{xire2}
\end{eqnarray}
where $K_m$ is the modified Bessel function of the second kind and a
prime on $K_m$ indicates a derivative with respect to its
appropriate argument. The coefficients $\alpha,\beta,\gamma$ and
$\varepsilon$ in Eqs. (\ref{xir}), (\ref{Pi01}), (\ref{Pi02}),
(\ref{Pe}) and (\ref{xire2}) are determined by the appropriate
boundary conditions.

%-----------------------------------------------------------------------------------------------
\section{Boundary conditions and dispersion relation}\label{dr}
The necessary boundary conditions at the perturbed tube boundary are
that the plasma displacement in the radial direction and the
magnetic pressure should be continuous as
\begin{eqnarray}
{\xi _{r{\rm i}} } \Big|_{r = a}  = {\xi _{r0} } \Big|_{r =
a},~~~~~~{\xi _{r0} } \Big|_{r = R} = {\xi _{r{\rm e}} } \Big|_{r =
R},\label{bc1}
\end{eqnarray}
\begin{eqnarray}
\begin{array}{l}
 {\delta P_{\rm i}  - \frac{{B_{\phi{\rm i} }^2}}{{4\pi a}}\xi _{r{\rm i}} } \Big|_{r = a}  = {\delta P_0  - \frac{{B_{\phi0}^2 }}{{4\pi a}}\xi _{r0} } \Big|_{r = a} ,\, \\
 \,{\delta P_0  - \frac{{B_{\phi0}^2 }}{{4\pi R}}\xi _{r0} } \Big|_{r = R}  = {\delta P_{\rm e} } \Big|_{r = R}. \label{bc2} \\
 \end{array}
\end{eqnarray}
Using the above boundary conditions and the solutions given by Eqs.
(\ref{xir}), (\ref{Pi01}), (\ref{Pi02}) for the internal and annulus
regions and Eqs. (\ref{Pe}), (\ref{xire2}) for the exterior region,
the dispersion relation is derived as
\begin{eqnarray}
\Big(\Xi_m^{0}\Pi_m^0-\Xi_m^{\rm i}\Xi_m^{\rm
e}\Big)\Big[1-(a/R)^{2m}\Big]~~~~~~ \nonumber\\
-\Xi_m^{\rm
i}\Big[\Xi_m^{0}-(a/R)^{2m}\Pi_m^0\Big]~~~~~~\nonumber\\+\Xi_m^{\rm
e}\Big[\Pi_m^0-(a/R)^{2m}\Xi_m^{0}\Big]
  = 0,\label{dp}
\end{eqnarray}
with
\begin{eqnarray}
\Xi_m^{\rm j}  = \frac{1}{m}\Big(\rho_{\rm j} \omega ^2  -
\frac{{B_0^2 k_z^2 }}{{4\pi}}\Big)~~~~~~~~~~~~~~~~\nonumber\\+
\frac{{A_{\rm j} }}{{4\pi m}}(2B_0 k_z + mA_{\rm j} )(1 -
m),\label{fmj}
\end{eqnarray}
\begin{eqnarray}
\Pi_m^0=- \frac{1}{m}\Big(\rho _0 \omega ^2  - \frac{{B_0^2 k_z^2
}}{{4\pi}}\Big)~~~~~~~~~~~~~\nonumber\\+ \frac{{A_0 }}{{4\pi
m}}(2B_0 k_z + mA_0 )(1 + m),\label{fm}
\end{eqnarray}
\begin{eqnarray}
\Xi_m^{\rm e}=\frac{B_0^2}{4\pi}\frac{k'^2}{k'R}\frac{K_m (k'R)}{K_m
^\prime (k'R)},~~~~~~~~~~~~~~~\label{fme}
\end{eqnarray}
where the superscript ${\rm j}$ in $\Xi_m^{\rm j}$ stands for ${\rm
i}$ and $0$ corresponding to the interior and annulus regions,
respectively.

Note that if we remove the annulus region by setting $a=R$ in the
dispersion relation (\ref{dp}) and using the thin flux tube
approximation for $K_m (x)\propto x^{-m}$ at small $x$, the result
yields
\begin{eqnarray}
\omega^2=C_k^2\Big\{k_z^2+\frac{A_{\rm
i}(m-1)}{2B_0^2}(2B_0k_z+A_{\rm i}m)\Big\},\label{omega}
\end{eqnarray}
where $C_k^2=\frac{B_0^2}{2\pi(\rho_{\rm{i}}+\rho_{\rm{e}})}$.
Equation (\ref{omega}) is same as Eq. (40) in Ruderman (2007).
The main result of Ruderman (2007) is that the twist does
not affect the kink modes in the particular case of having
$B_\phi\propto r$ and Eq. (\ref{omega}) shows that we get the same
frequencies as in the case that $A_{\rm i}=0$. This result also has
been already obtained by Goossens, Hollweg $\&$ Sakurai (1992). Note
that Eq. (\ref{dp}) shows that even in the presence of annulus, the
internal twist does not affect the kink $(m=1)$ modes. Because the
internal twist, $A_{\rm{i}}$, only appears in Eq. (\ref{fmj}) and
when $m=1$ then it has no contribution.

In subsections 3.1 and 3.2, we show that the dispersion relation
(\ref{dp}) for the cases $A_{\rm i}=0$ and $A_0=0$, respectively,
reduces to two known cases. Also in subsection 3.3, we give an
analytical solution for the dispersion relation (\ref{dp}).

%-----------------------------------------------------------------------------------------------
\subsection{Case $A_{\rm i}=0$}

Here we show that the dispersion relation (\ref{dp}) in the absence
of internal twist, i.e. $A_{\rm i}=0$, can be obtained from the
dispersion relation, Eq. (6b), given by Carter $\&$ Erd\'{e}lyi
(2008) under the thin tube (TT) approximation (or long-wavelength
limit), i.e. $k_za\ll 1$. The dispersion relation Eq. (6b) in Carter
$\&$ Erd\'{e}lyi (2008) for body waves has the form
\begin{eqnarray}
\frac{\Xi_{aY}-\Xi_{\rm i}+\Xi_{aY}\Xi_{\rm
i}\frac{A_0^2}{4\pi}}{\Xi_{aJ}-\Xi_{\rm i}+\Xi_{aJ}\Xi_{\rm
i}\frac{A_0^2}{4\pi}}\frac{Y_m(n_0a)}{J_m(n_0a)}=~~~~~~~~~~~~~~~\nonumber\\\frac{Y_m(n_0R)}{J_m(n_0R)}\frac{\Xi_{RY}-\Xi_{\rm
e}+\Xi_{RY}\Xi_{\rm e}\frac{A_0^2}{4\pi}}{\Xi_{RJ}-\Xi_{\rm
e}+\Xi_{RJ}\Xi_{\rm e}\frac{A_0^2}{4\pi}},\label{6b}
\end{eqnarray}
which is valid for thick magnetic tubes with twisted annulus in the
incompressible limit. Note that according to Edwin \& Roberts (1983)
there are no surface waves in zero-beta approximation which is
compatible with coronal conditions. Although the plasma in our model
is compressible, the results concerning the kink ($m=1$) modes for
incompressible plasmas given by Carter $\&$ Erd\'{e}lyi (2008) can
be applied to coronal loops despite that the coronal plasma is a
low-beta plasma (see Carter \& Erd\'{e}lyi 2007; Erd\'{e}lyi \&
Fedun 2007). Recently, Goossens et al. (2009) showed that in the TT
approximation neglecting contributions proportional to $(k_za)^2$
then the frequencies of the kink wave are the same in the three
cases including a compressible pressureless plasma, an
incompressible plasma and a compressible plasma which allows for MHD
radiation.

Under the TT approximation, we have
\begin{equation}
\frac{{xK_m '(x)}}{{K_m (x)}} = \frac{{xY_m '(x)}}{{Y_m (x)}}
=-m+O(x^2),\label{KY}
\end{equation}
\begin{equation}
\frac{{xI_m '(x)}}{{I_m (x)}} = \frac{{xJ_m '(x)}}{{J_m (x)}} =
m+O\big(x^2\big),\label{IJ}
\end{equation}
where ($J_m,Y_m$) and ($I_m,K_m$) are the Bessel and modified Bessel
functions of the first and second kind, respectively. Using Eqs.
(\ref{KY}), (\ref{IJ}), the different terms appeared in Eq.
(\ref{6b}) given by Carter \& Erd\'{e}lyi (2008) reduce to
\begin{equation}
\Xi _{aY}  = \Xi _{RY}= - \frac{m}{{\rho _0 }}\frac{1}{{\big(\omega
^2 - \omega_{A_0}^2 \big) - \frac{{2A_0 \omega _{A_0} }}{{\sqrt
{4\pi \rho _0 } }}}},\label{XiaY}
\end{equation}
\begin{equation}
\Xi_{aJ}=\Xi_{RJ}=\frac{m}{\rho_0}\frac{1}{{\big(\omega ^2 - \omega
_{A_0}^2 \big) + \frac{{2A_0 \omega_{A_0} }}{{\sqrt {4\pi \rho _0 }
}}}},\label{XiaJ}
\end{equation}
\begin{equation}
\Xi_{\rm i}  = \frac{m}{{\rho_{\rm i} \big(\omega ^2  -
\omega_{A_{\rm i}}^2 \big)}},\label{Xii}
\end{equation}
\begin{equation}
\Xi_{\rm e}  = \frac{{ - m}}{{\rho_{\rm e} \big(\omega ^2  -
\omega_{A_{\rm e}}^2\big)}},\label{Xie}
\end{equation}
where following Carter \& Erd\'{e}lyi (2008), $\omega_{A_0}$,
$\omega_{A_{\rm i}}$, and $\omega_{A_{\rm e}}$ are the Alfv\'{e}n
frequencies in the annulus, internal and external regions,
respectively, given by
\begin{equation}
\omega_{A_0}=\frac{1}{\sqrt{4\pi\rho_0}}(mA_0+k_zB_0),\label{wA0}
\end{equation}
\begin{equation}
\omega_{A_{\rm i}}=\frac{k_zB_{\rm i}}{\sqrt{4\pi\rho_{\rm i}}},
\end{equation}
\begin{equation}
\omega_{A_{\rm e}}=\frac{k_zB_{\rm e}}{\sqrt{4\pi\rho_{\rm
e}}}.\label{wAe}
\end{equation}
Using Eqs. (\ref{fmj}) to (\ref{fme}) and Eqs. (\ref{KY}) to
(\ref{wAe}), one can rewrite the terms $\Xi
_{Y}:=\Xi_{aY}=\Xi_{RY}$, $\Xi _{J}:=\Xi_{aJ}=\Xi_{RJ}$, $\Xi_{\rm
i}$, and $\Xi_{\rm e}$ appeared in the dispersion relation
(\ref{6b}) as follows
\begin{equation}
\Xi_{\rm i}  = \frac{1}{{\Xi _m^{\rm i} }},\label{xiRE}
\end{equation}
\begin{equation}
\Xi _J  = \frac{1}{{\Xi_m^0+\frac{{A_0^2 }}{{4\pi }}}},
\end{equation}
\begin{equation}
\Xi _Y = \frac{1}{{\Pi_m^0+\frac{{A_0^2 }}{{4\pi }}}},
\end{equation}
\begin{equation}
\Xi_{\rm e}=\frac{4\pi m}{B_0^2k'^2}=-\frac{1}{\Xi_m^{\rm e}}.
\end{equation}
Also under the TT approximation we have
\begin{equation}
\frac{Y_m(n_0R)}{Y_m(n_0a)}\frac{J_m(n_0a)}{J_m(n_0R)}=\Big(\frac{a}{R}\Big)^{2m}.\label{aR}
\end{equation}
One can easily show that substituting Eqs. (\ref{xiRE}) to
(\ref{aR}) into (\ref{6b}) yields
\begin{equation}
R^{2m}\frac{\Xi_m^{0}+\Xi_m^{\rm e}}{\Pi_m^0+\Xi_m^{\rm
e}}=a^{2m}\frac{\Xi_m^{0}-\Xi_m^{\rm i}}{\Pi_m^0-\Xi_m^{\rm
i}},\label{dp1}
\end{equation}
which is nothing but Eq. (\ref{dp}) in which the terms $a^{2m}$ and
$R^{2m}$ have been grouped.

%-----------------------------------------------------------------------------------------------
\subsection{Case $A_{0}=0$}
Here we present that the dispersion relation (\ref{dp1}) in the
absence of twist in the annulus region, i.e. $A_{0}=0$, is same as
the dispersion relation Eq. (11) in Carter \& Erd\'{e}lyi (2007) for
the kink ($m=1$) modes in the TT approximation. By setting
$A_{0}=0$, Eqs. (\ref{fmj}) to (\ref{fme}) reduce to
\begin{equation}
\Xi_1^{\rm i}=\rho_{\rm i}\big(\omega^2-\omega_{A_{\rm
i}}^2\big),\label{xi1i}
\end{equation}
\begin{equation}
\Xi_1^{0}=\rho_{0}\big(\omega^2-\omega_{A_{0z}}^2\big),\label{xi10}
\end{equation}
\begin{equation}
\Pi_1^0=-\rho_{0}\big(\omega^2-\omega_{A_{0z}}^2\big),\label{xi1}
\end{equation}
\begin{equation}
\Xi_1^{\rm e}=\rho_{\rm e}\big(\omega^2-\omega_{A_{\rm
e}}^2\big),\label{xi1e}
\end{equation}
where
\begin{equation}
\omega_{A_{0z}}=\frac{k_zB_0}{\sqrt{4\pi\rho_0}}\label{wA0z}.
\end{equation}
With the help of above expressions, Eq. (\ref{dp1}) yields
\begin{eqnarray}
\frac{{Q_0^{\rm i}  + 1}}{{Q_0^{\rm i}  - 1}} -
\Big(\frac{a}{R}\Big)^{2} \frac{{Q_0^{\rm e} - 1}}{{Q_0^{\rm e}  +
1}}= 0,\label{dpQ0}
\end{eqnarray}
where
\begin{equation}
Q_0^{\rm i}  = \frac{{\rho _{\rm i} }}{{\rho _0 }}\left(\frac{\omega
^2 - \omega _{A_{\rm i}}^2}{\omega ^2  - \omega _{A_{0z}}^2}\right),
\end{equation}
\begin{equation}
Q_0^{\rm e}  = \frac{\rho _{\rm e}}{\rho _0}\left(\frac{\omega ^2 -
\omega _{A_{\rm e}}^2}{\omega ^2  - \omega _{A_{0z}}^2}\right).
\end{equation}
Equation (\ref{dpQ0}) is same as the dispersion relation (11) in
Carter \& Erd\'{e}lyi (2007) for the kink ($m=1$) modes in the TT
approximation.

%-----------------------------------------------------------------------------------------------
\subsection{Analytical solution of the dispersion relation}
Here we try to solve the dispersion relation (\ref{dp}),
analytically. To do this we apply the TT approximation, Eq.
(\ref{KY}), to the expression $\Xi_m^{\rm e}$, Eq. (\ref{fme}). This
reduces the dispersion relation (\ref{dp}) to a second order
equation in terms of $\omega^2$  which can now be solved
analytically. To write the solutions in a simple form, we define two
expressions $X_m^{\rm j}$ and $Z_m^0$ as follows
\begin{equation}
X_m^{\rm j}=\frac{A_{\rm j}}{B_0k_z}\left(2+\frac{mA_{\rm
j}}{B_0k_z}\right)(1-m),\label{X}
\end{equation}
\begin{equation}
Z_m^0=\frac{A_0}{B_0k_z}\left(2+\frac{mA_0}{B_0k_z}\right)(1+m),\label{Z}
\end{equation}
which are proportional to the second terms appeared in Eqs.
(\ref{fmj}) and (\ref{fm}), respectively. These expressions contain
all twist parameters in the dispersion relation.

Using Eqs. (\ref{X}) and (\ref{Z}), one can rewrite Eq.
(\ref{dp}) as
\begin{equation}
c_m\left(\frac{\sqrt{4\pi\rho_{\rm
i}}}{B_0k_z}\right)^4\omega^4-c_{ml}\left(\frac{\sqrt{4\pi\rho_{\rm
i}}}{B_0k_z}\right)^2\omega^2+\tilde{c}_{ml}=0,\label{analy}
\end{equation}
where $c_m$, $c_{ml}$ and $\tilde{c}_{ml}$ are constants defined as
\begin{eqnarray}
c_m=\left[\frac{\rho_{\rm e}}{\rho_{\rm
i}}+\left(\frac{\rho_0}{\rho_{\rm
i}}\right)^2\right]\left(1-(a/R)^{2m}\right)
+\frac{\rho_0}{\rho_{\rm i}}\left(1+\frac{\rho_{\rm e}}{\rho_{\rm
i}}\right)\left(1+(a/R)^{2m}\right),\label{cm}
\end{eqnarray}
\begin{eqnarray}
c_{ml}=2\left(1+\frac{\rho_{\rm e}}{\rho_{\rm
i}}+2\frac{\rho_0}{\rho_{\rm i}}\right)
+\left[\frac{\rho_0}{\rho_{\rm i}}+\frac{\rho_{\rm e}}{\rho_{\rm
i}}+\left(1-\frac{\rho_0}{\rho_{\rm
i}}\right)(a/R)^{2m}\right]Z_m^0~~~~~~~~~~~~~~~~~~~~~~~~\nonumber\\
-\left[1+\frac{\rho_0}{\rho_{\rm i}}+\left(\frac{\rho_{\rm
e}}{\rho_{\rm i}}-\frac{\rho_0}{\rho_{\rm
i}}\right)(a/R)^{2m}\right]X_m^0 -\left[\frac{\rho_0}{\rho_{\rm
i}}+\frac{\rho_{\rm e}}{\rho_{\rm i}}+\left(\frac{\rho_0}{\rho_{\rm
i}}-\frac{\rho_{\rm e}}{\rho_{\rm
i}}\right)(a/R)^{2m}\right]X_m^{\rm i},\label{cml}
\end{eqnarray}
\begin{eqnarray}
\tilde{c}_{ml}=4-2\Big(X_m^{\rm
i}+X_m^0-Z_m^0\Big)-\Big(Z_m^0-X_m^{\rm
i}\Big)X_m^0-(a/R)^{2m}\Big(X_m^{\rm
i}-X_m^0\Big)Z_m^0.\label{ctilda}
\end{eqnarray}
Here $c_m$ depends only on the azimuthal mode number $m$ but
$c_{ml}$ and $\tilde{c}_{ml}$ contain both the azimuthal $m$ and
longitudinal $l$ mode numbers. Note that $l$ appears in $k_z=l\pi
/L$.

Now solving Eq. (\ref{analy}) gives the eigenfrequencies
\begin{equation}
\omega_{nml}=\frac{k_zB_0}{\sqrt{4\pi\rho_i}}\left(\frac{c_{ml}\pm\sqrt{c_{ml}^2-4\tilde{c}_{ml}c_{m}}}{2c_m}~\right)^{1/2},\label{analysol}
\end{equation}
where the subscript $n$ (or $\pm$ signs) denotes the radial mode
number corresponding to the lower $(n=1)$ and upper $(n=2)$
frequencies of the two nonaxisymmetric modes. Note that for kink
($m=1$) modes from Eq. (\ref{X}) we get $X_m^{\rm i}=X_m^{0}=0$.
Also from Eqs. (\ref{cml}) and (\ref{ctilda}) for $m=1$ both
$c_{ml}$ and $\tilde{c}_{ml}$ are positive. On the other hand $c_m$
is always positive. Therefore, from Eq. (\ref{analysol}) the $-$ and
$+$ signs are corresponding to the radial mode numbers $n=1$ and
$n=2$, respectively. Since our model is based on the TT
approximation, the numerical solution of the dispersion relation
(\ref{dp}) in Section \ref{NR} is approximately same as the
analytical solution (\ref{analysol}).

Note that in addition to the solutions (\ref{analysol}), there is a
set of broadband body modes which has been already predicted by
Carter $\&$ Erd\'{e}lyi (2008). However, these modes are lost from
the solutions of the dispersion relation (\ref{6b}) under the TT
approximation. Since our model is based on the TT approximation,
therefore the infinite set of body modes are absent in our
calculations. In accordance with the classification introduced by
Roberts (1981), the modes both described and not described by Eq.
(\ref{dp}) are body waves in the zero-beta plasma approximation. To
distinguish the kink mode described by Eq. (\ref{dp}) from those not
described by Eq. (\ref{dp}), Ruderman and Roberts (2002) suggested
calling the solutions (\ref{analysol}) ``global kink modes,''
retaining the name ``body kink modes'' for all other kink modes. We
generalize this convention for fluting modes and call fluting modes
described by Eq. (\ref{dp}) ``global fluting modes,'' retaining the
name ``body fluting modes'' for all other fluting modes.

%-----------------------------------------------------------------------------------------------
\section{Numerical results}\label{NR}
To solve the dispersion relation (\ref{dp}), numerically, we
choose the physical parameters $L=10^5$ km, $a/L=0.01$, $\rho_{{\rm
e}}/\rho_{{\rm i}}=0.1$, $\rho_{0}/\rho_{{\rm i}}=0.5$, $\rho_{{\rm
i}}=2\times 10^{-14}$ g cm$^{-3}$, $B_{0}=100$ G. For such a loop
one finds $v_{A_{\rm i}}=\frac{B_0}{\sqrt{4\pi\rho_{{\rm i}}}}=2000$
km s$^{-1}$, $\omega_{A_{\rm i}}:=\frac{v_{A_{\rm i}}}{{\rm
L}}=0.02$ rad s$^{-1}$. In what follows, we illustrate our
numerical studies in the two separate equilibrium cases where there
is (i) Twist in both the core and annulus regions (ii) Twist in the
core and no twist in the annulus.

%-----------------------------------------------------------------------------------------------
\subsection{Case $A_{\rm i}\neq0$ and $A_0\neq0$}
\begin{figure}
\includegraphics{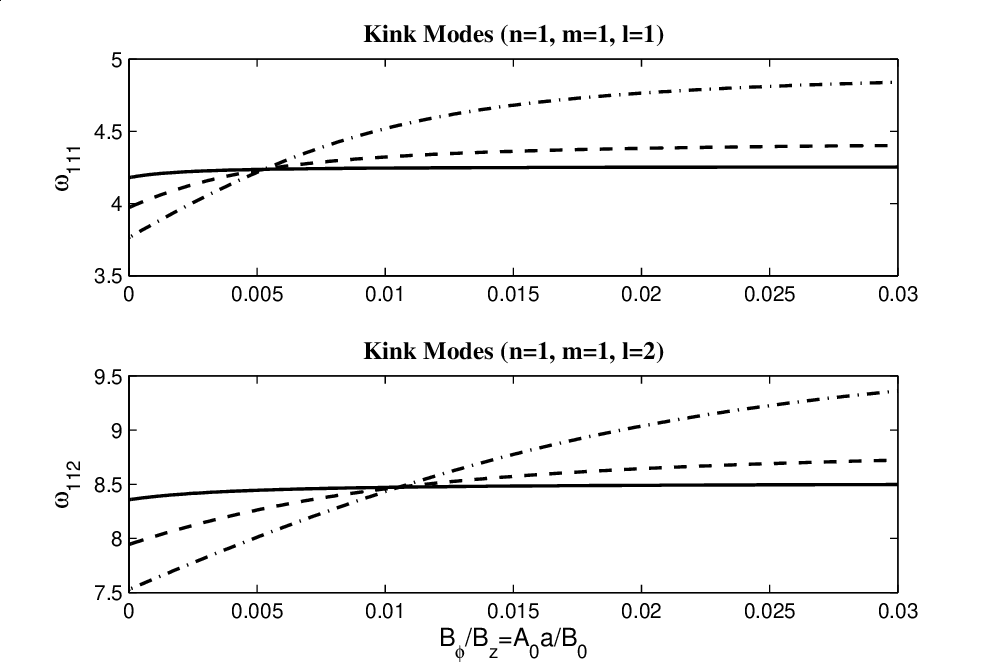}
      \vspace{6cm}
      \caption[]{Frequencies of the fundamental and its first overtone kink $(m=1)$ modes with radial mode number $n=1$
      versus the twist parameter of the annulus, $B_{\phi}/B_z=\frac{A_0a}{B_0}$, for different
relative core widths $a/R=$ 0.65 (dash-dotted), 0.9
(dashed) and 0.99 (solid). The loop parameters are: $L=10^5$ km,
$a/L=0.01$, $\rho_{{\rm e}}/\rho_{{\rm i}}=0.1$,
$\rho_{0}/\rho_{{\rm i}}=0.5$, $\rho_{{\rm i}}=2\times 10^{-14}$ g
cm$^{-3}$, $B_{0}=100$ G. Frequencies are in units of the interior
Alfv\'{e}n frequency, $\omega_{\rm A_i}= 0.02{\rm~rad~s^{-1}}$.}
         \label{w1l1-w112}
   \end{figure}

\begin{figure}
\includegraphics{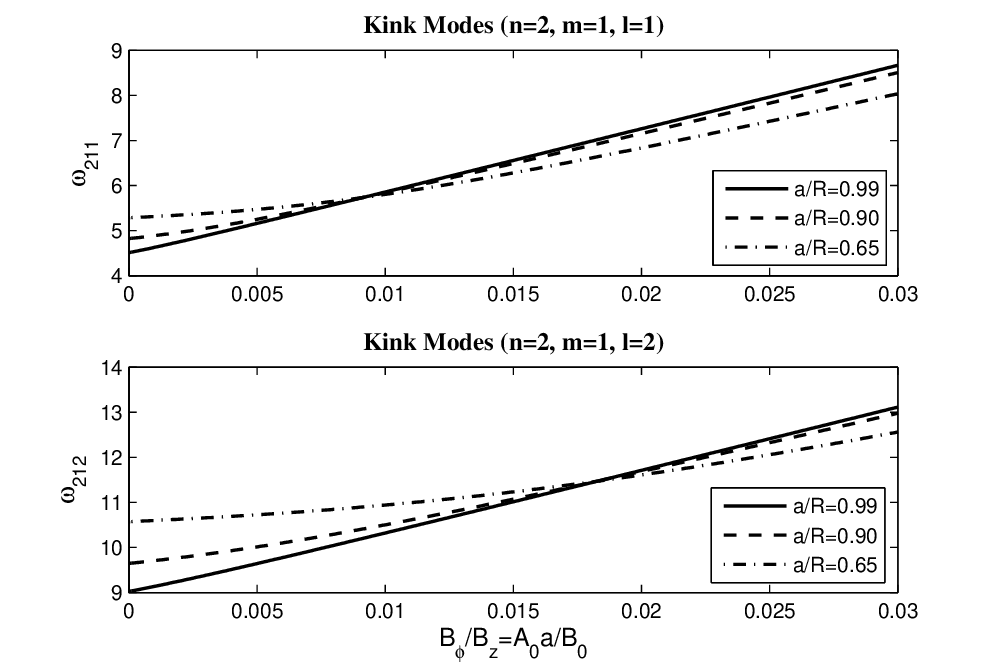}
      \vspace{4.5cm}
      \caption[]{Same as Fig. \ref{w1l1-w112}, for the kink ($m=1$) modes with radial mode number $n=2$.}
         \label{w2l1-w212}
   \end{figure}

The effect of twisted magnetic field on the frequencies $\omega$ is
calculated by the numerical solution of the dispersion relation, Eq.
(\ref{dp}). Figures \ref{w1l1-w112} to \ref{w2l1-w212} show the
frequencies of the fundamental and first overtone $l=1,2$ kink $(m =
1)$ modes with radial mode numbers $n=1,2$ versus the twist
parameter of the annulus, $B_{\phi}/B_z:=\frac{A_0a}{B_0}$, and for
different relative core widths $a/R=(0.65, 0.9, 0.99)$. Note that
here the parameter $A_{\rm i}$ does not need to be set explicitly.
Because the second term in Eq. (\ref{fmj}) containing the
contribution of the parameter $A_{\rm i}$ is automatically removed
for the kink $(m=1)$ modes.

Figures \ref{w1l1-w112} to \ref{w2l1-w212} reveal that: i) for a
given $a/R$, the frequencies increase when the twist parameter of
the annulus increases. The result is in good agreement with that
obtained by Carter $\&$ Erd\'{e}lyi (2008) and Karami $\&$ Barin
(2009). ii) For a given $n$ and $a/R$, when the longitudinal mode
number, $l$, increases, the frequencies increase. iii) For a given
$l$, $a/R$ and $B_{\phi}/B_z$, when the radial mode number, $n$,
increases, the frequencies increase. iv) For $n=1$, when $a/R$ goes
to unity then the frequencies become independent of $B_{\phi}/B_z$.
Therefore in the absence of the annulus, the twist does not affect
the kink modes in the specific case of having $B_\phi\propto r$.
This is in good agreement with that obtained by Goossens, Hollweg
$\&$ Sakurai (1992) and Ruderman (2007). v) For $n=2$, when $a/R$
goes to unity exactly then the second mode is removed. This is
expected to be occurred because for $a/R=1$ we have only one
boundary in the tube corresponding to one mode.

\begin{figure}
\includegraphics{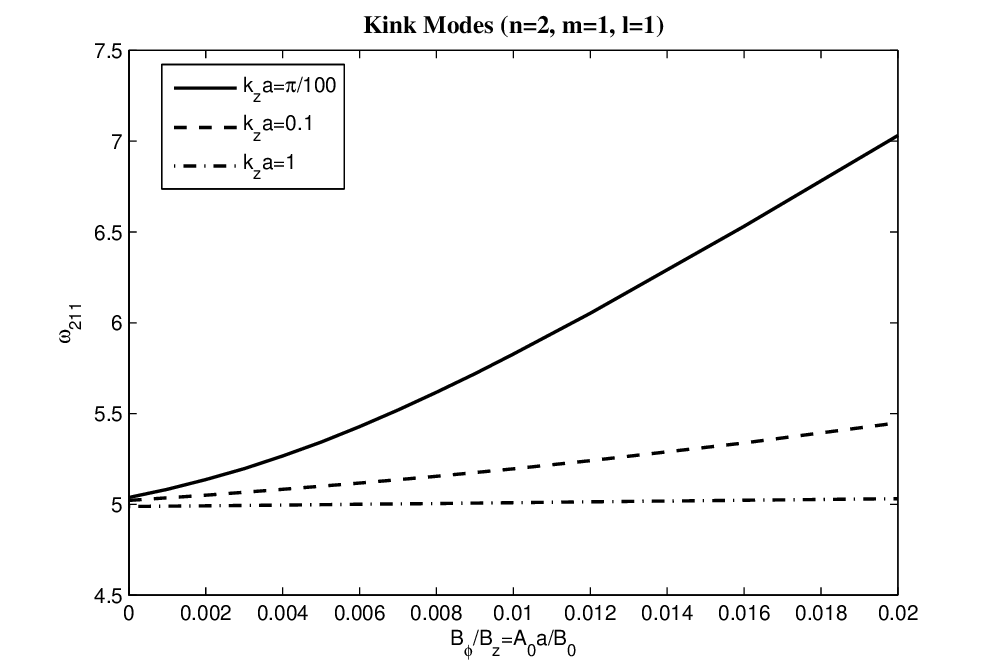}
      \vspace{5.8cm}
      \caption[]{Frequencies of the kink
$(m=1)$ modes with radial mode number $n=2$ versus the twist
parameter of the annulus, $B_{\phi}/B_z=\frac{A_0a}{B_0}$, for
different values of $k_za=1$ (dash-dotted), 0.1 (dashed) and
$\pi/100$ (solid). Here $a/R=0.8$ and other auxiliary parameters as
in Fig. \ref{w1l1-w112}.} \label{w111dif-ator}
\end{figure}

To compare our results with those of Carter \& Erd\'{e}lyi (2008),
we use Eq. (\ref{6b}) and obtain the frequencies of the kink $(m =
1)$ modes with radial mode number $(n=2)$ versus the twist parameter
of the annulus and for different values of $k_za=\pi
a/L=(\pi/100,0.1,1)$. The results are displayed in Fig.
\ref{w111dif-ator}. The result for $k_za=1$ has been already
obtained by Carter $\&$ Erd\'{e}lyi (2008) for incompressible flux
tube with twisted annulus. Figure \ref{w111dif-ator} shows that for
a slender tube with $k_za=\pi/100$ there is a much greater variation
with the twist in the annulus than what Carter $\&$ Erd\'{e}lyi
(2008) found for thicker tube with $k_za=1$.

One important problem about the frequencies displayed in Figs.
\ref{w1l1-w112} and \ref{w2l1-w212} is that in all four panels in
these figures, all the curves for different values of $a/R$
intersect at a single point. For instance, in Fig.
\ref{w1l1-w112} for the fundamental and first overtone kink
modes, the frequencies of the intersection points are
$\omega_{111}=4.23$ and $\omega_{112}=2\omega_{111}=8.46$,
respectively. Also the location of the intersection point for the
fundamental kink mode $\omega_{111}$ occurs at
$B_{\phi}/B_z=\frac{A_0a}{B_0}=0.0055$ and for the first overtone
kink mode $\omega_{112}$ the intersection occurs exactly at twice
the value as for the fundamental mode. The frequency and the twist
of the intersection point can be obtained from the dispersion
relation. To do this we rewrite the dispersion relation (\ref{dp1})
for the kink ($m=1$) modes as
\begin{equation} f-\Big(\frac{a}{R}\Big)^{2}g=0,\label{dp2}
\end{equation}
where
\begin{equation}
f=\frac{\Pi_1^0-\Xi_1^{\rm i}}{\Xi_1^0-\Xi_1^{\rm i}},\label{ff}
\end{equation}
\begin{equation}
g=\frac{\Pi_1^0+\Xi_1^{\rm e}}{\Xi_1^0+\Xi_1^{\rm e}}.\label{gg}
\end{equation}
If we set $f=g=0$, then the result of the dispersion relation
(\ref{dp2}) would be independent of the term $a/R$. Therefore the
set of equations $\Pi_1^0-\Xi_1^{\rm i}=0$ and $\Pi_1^0+\Xi_1^{\rm
e}=0$ by the help of Eqs. (\ref{fmj}) to (\ref{fme}) in the TT
approximation yield
\begin{equation} \omega =C_kk_z= \frac{B_0}{\sqrt{2\pi(\rho_{\rm i} +
\rho_{\rm e})}}\frac{l\pi}{L},\label{wcut}
\end{equation}
\begin{equation}
\frac{B_{\phi}}{B_z}=\frac{A_0a}{B_0}=\left(\sqrt{\frac{\rho_{\rm
i}+\rho_0}{\rho_{\rm i}+\rho_{\rm e}}}-1\right)\frac{l\pi
a}{L},\label{twistcut}
\end{equation}
which show that the values of both the frequency and
the twist of the intersection point depend on the
wavelength as observed. Taking $L=10^5$ km, $a/L=0.01$, $\rho_{{\rm
e}}/\rho_{{\rm i}}=0.1$, $\rho_{0}/\rho_{{\rm i}}=0.5$, and
$B_{0}=100$ G, then Eqs. (\ref{wcut}) and (\ref{twistcut}) give
$\omega/\omega_{A_{\rm i}}=4.24~l$ and
$B_{\phi}/B_z=\frac{A_0a}{B_0}=0.0053~l$ which are in good agreement
with the numerical results obtained for the frequency and the twist
of the intersection point in Fig. \ref{w1l1-w112}.

\begin{figure}
\includegraphics{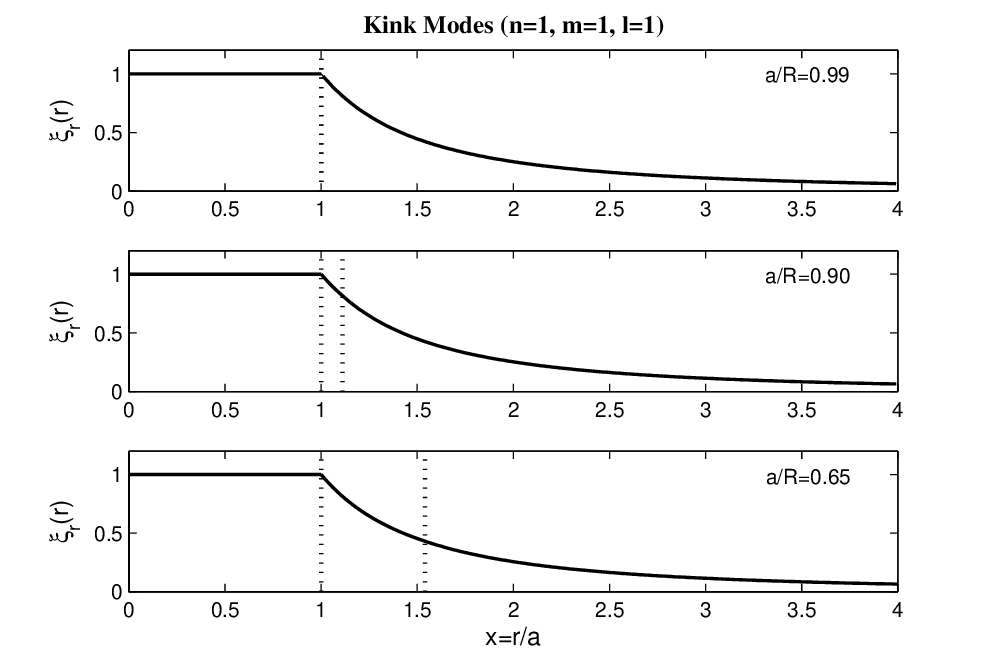}
      \vspace{5.5cm}
\caption[] {Radial component of the eigenfunctions of the
fundamental kink $(m=1)$ modes with radial mode number $n=1$ against
fractional radius $x=r/a$ for $B_{\phi}/B_z=\frac{A_0a}{B_0}=0.0055$
and different relative core widths $a/R=$ 0.65, 0.9 and 0.99.}
         \label{eigen1}
   \end{figure}

\begin{figure}
\includegraphics{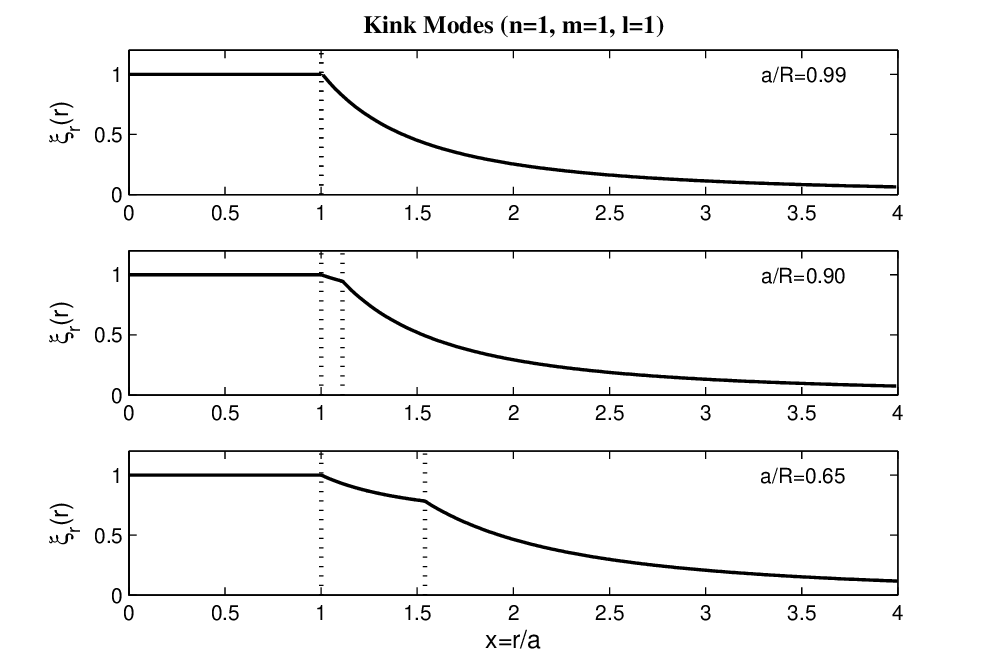}
      \vspace{6.7cm}
\caption[] {Same as Fig. \ref{eigen1}, for
$B_{\phi}/B_z=\frac{A_0a}{B_0}=0.02$.}
         \label{eigen2}
   \end{figure}

To investigate this problem from another point of view, the radial
component of the fundamental kink eigenfunctions, $\xi_r(r)$, is
studied. To do this, using the eigenfrequencies obtained from the
dispersion relation (\ref{dp}) and applying the boundary conditions
(\ref{bc1}) and (\ref{bc2}) to the solutions (\ref{xir}),
(\ref{Pi01}), (\ref{Pi02}), (\ref{Pe}) and (\ref{xire2}) one can
obtain the coefficients $\alpha$, $\beta$, $\gamma$ and
$\varepsilon$. Then with the help of Eqs. (\ref{xir}) and
(\ref{xire2}) the radial component of the fundamental kink
eigenfunctions $\xi_r(r)$ can be obtained. The results of $\xi_r(r)$
for different relative core widths $a/R=(0.65, 0.9, 0.99)$ and for
two different values of the twist parameter of the annulus
$B_{\phi}/B_z=\frac{A_0a}{B_0}=0.0055$ and $0.02$ are plotted in
Figs. \ref{eigen1} and \ref{eigen2}, respectively. Note that the
intersection occurs at $B_{\phi}/B_z=0.0055$. Comparing Fig.
\ref{eigen1} with \ref{eigen2} we find out for the twist parameter
of the annulus $0.0055$, the eigenfunctions at the second boundary
behave smoothly. This means that at the intersection point, the
location of the second boundary is not important or physically the
variation of the thickness of the annulus does not affect the
eigenfunctions or eigenfrequencies. Figure \ref{eigen2} clears that
for the twist parameter of the annulus $0.02$ where the intersection
does not occur, the behavior of the eigenfunctions depends on the
thickness of the annulus region.

\begin{figure}
\includegraphics{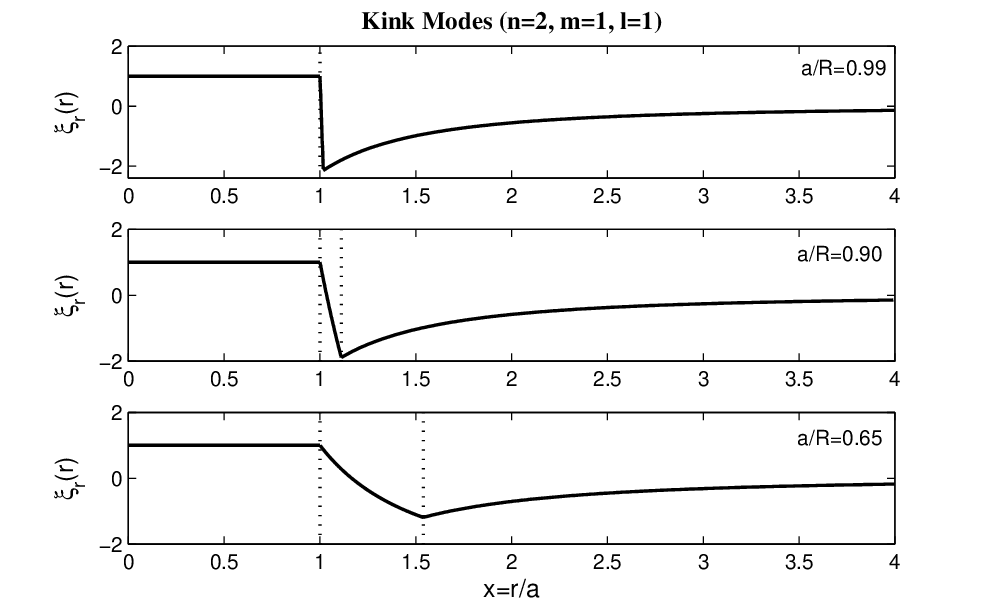}
      \vspace{6.7cm}
\caption[] {Radial component of the eigenfunctions of the
fundamental kink $(m=1)$ modes with radial mode number $n=2$ against
fractional radius $x=r/a$ for $B_{\phi}/B_z=\frac{A_0a}{B_0}=0.02$
and different relative core widths $a/R=$ 0.65, 0.9 and 0.99.}
         \label{eigen3}
   \end{figure}

To compare the kink oscillations for different radial mode numbers
$n$, the radial component of the kink eigenfunctions with $n=2$ is
plotted in Fig. \ref{eigen3} for the twist parameter of the annulus
$B_{\phi}/B_z=\frac{A_0a}{B_0}=0.02$ and for different relative core
widths $a/R=(0.65, 0.9, 0.99)$. \textbf{Figure \ref{eigen3} shows
that contrary to the $n=1$ mode in which the core and annulus
regions oscillate with the same phase (see Fig. \ref{eigen2}), for
$n=2$ the kink oscillations of the core and annulus region, having
opposite signs of $\xi_r(r)$, are out of phase with each other. This
is in agreement with the previous results obtained by Mikhlyaev \&
Solov'ev (2005) and Ruderman \& Erd\'{e}lyi (2009).}

\begin{figure}
\includegraphics{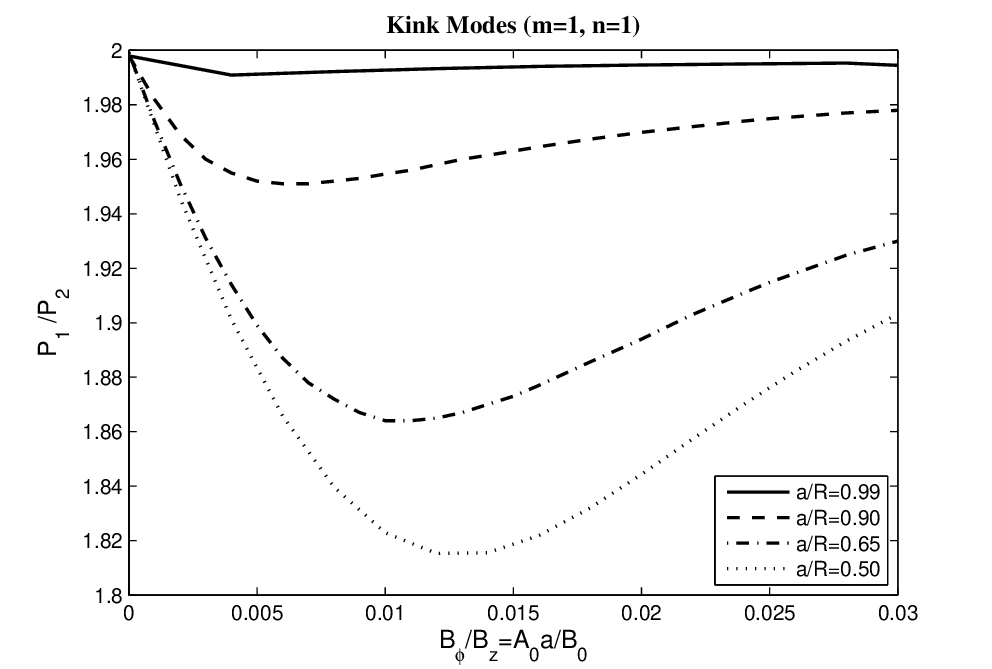}
      \vspace{6.0cm}
\caption[]{The period ratio $P_1/P_2$ of the fundamental and its
first overtone kink $(m=1)$ modes with radial mode number $n=1$
versus the twist parameter of the annulus for different relative
core widths $a/R=$ 0.5 (dotted), 0.65 (dash-dotted), 0.9 (dashed)
and 0.99 (solid). Auxiliary parameters as in Fig. \ref{w1l1-w112}.}
         \label{n1-p1p2}
   \end{figure}

\begin{figure}
\includegraphics{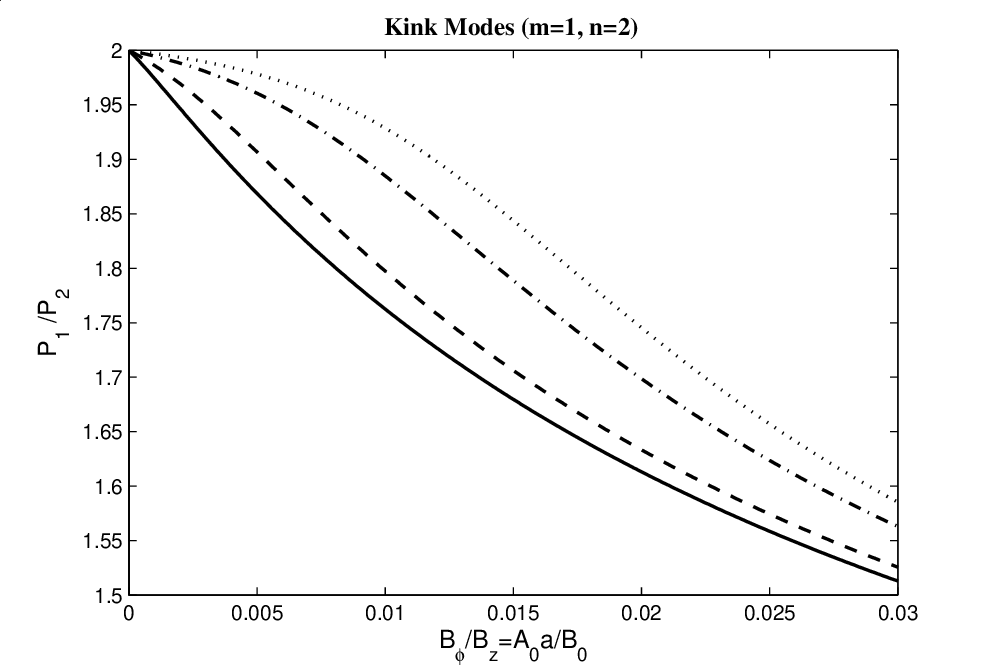}
      \vspace{7.0cm}
\caption[] {Same as Fig. \ref{n1-p1p2}, for the kink ($m=1$) modes
with radial mode number $n=2$.}
         \label{n2-p1p2}
   \end{figure}

The period ratio $P_1/P_2$ of the fundamental and first overtone,
$l=1,2$ modes of the kink $(m=1)$ waves with $n=1,2$ versus the
twist parameter of the annulus are plotted in Figs. \ref{n1-p1p2} to
\ref{n2-p1p2}. Figures \ref{n1-p1p2} to \ref{n2-p1p2} show that: i)
the period ratio $P_1/P_2$ with increasing the twist parameter of
the annulus, for $n=1$ decreases from 2 (for untwisted loop), comes
down to a minimum and then increases. Whereas for $n=2$, it
decreases from 2 and approaches below 1.6 for $a/R=0.5$, for
instance. Note that when the twist is zero, the diagrams of
$P_1/P_2$ do not start exactly from 2. This may be caused by the
radial structuring $(\rho_0\neq\rho_i$, $\rho_e\neq\rho_i)$. But for
the selected thin tube with $a/L=0.01$, this departure is very
small, $O(10^{-4})$, and doesn't show itself in the diagrams (see
McEwan et al. 2006). ii) For a given $B_{\phi}/B_z$, the period
ratio $P_1/P_2$ for $n=1$ increases and for $n=2$ decreases when the
relative core width increases. Figure \ref{n1-p1p2} clears that for
the kink modes $(m=1,n=1)$ with $a/R=0.5$, for both
$B_{\phi}/B_z=0.0107$ and 0.0153 the ratio $P_1/P_2$ is 1.82. This
is in good agreement with the period ratio observed by Van
Doorsselaere, Nakariakov $\&$ Verwichte (2007), $1.82\pm0.08$,
deduced from the observations of TRACE. See also McEwan, D\'{i}az
$\&$ Roberts (2008). For more observational examples of the period
ratio, we estimate the twist parameter of the annulus for the kink
modes $(m=1,n=1)$ with different relative core widths. The results
which can give the same $P_1/P_2$ observed by the TRACE are
summarized in Table \ref{table}.

Note that the results of Fig. \ref{n1-p1p2} and Table \ref{table}
show that the period ratio $P_1/P_2$ of the kink $(m=1)$ modes for
$n=1$ is not a monotonic function of the twist parameter of the
annulus. The analytical solution (\ref{analysol}) also confirms this
behavior. Therefore, we conclude that the value of the twist
parameter in coronal loops cannot be determined uniquely using the
model studied here.

Finally, it is worth to say some remarks regarding the coronal
seismology using the period ratio $P_1/P_2$ of the kink ($m=1$)
modes. Although the kink modes with $n=1$ and $n=2$ can be excited
in the flux tube, to compare the results with the observation, we
considered only the period ratio $P_1/P_2$ of the fundamental
($l=1$) and its first overtone ($l=2$) kink $(m=1)$ modes with
$n=1$. Because as we already mentioned in the case $n=2$ the core
and annulus oscillate with the opposite phases. This yields small
transverse global displacement for the tube which is not compatible
with the observed displacements of coronal loops, roughly 20 Mm from
the loop top (see Verwichte et al. 2004). The other possibility for
explaining the observed $P_1/P_2$ is the longitudinal fundamental
($l=1$) kink modes of $n=1$ and $n=2$. But this case also cannot
justify the observations. Following Verwichte et al. 2004, the
long-period oscillation is the fundamental mode, with a maximum
amplitude at the loop top. The short period oscillation is its
second harmonic. It has a node at the loop top, i.e. $l=2$. One
notes that besides the two global kink modes with $n=1,2$, an
infinite set of body kink modes can also be exited in the tube.
However, the body modes due to having oscillatory displacements
inside the tube cannot drive the loop to oscillate globally.

%-----------------------------------------------------------------------------------------------
\subsection{Case $A_{\rm i}\neq0$ and $A_0=0$}
\begin{figure}
\includegraphics{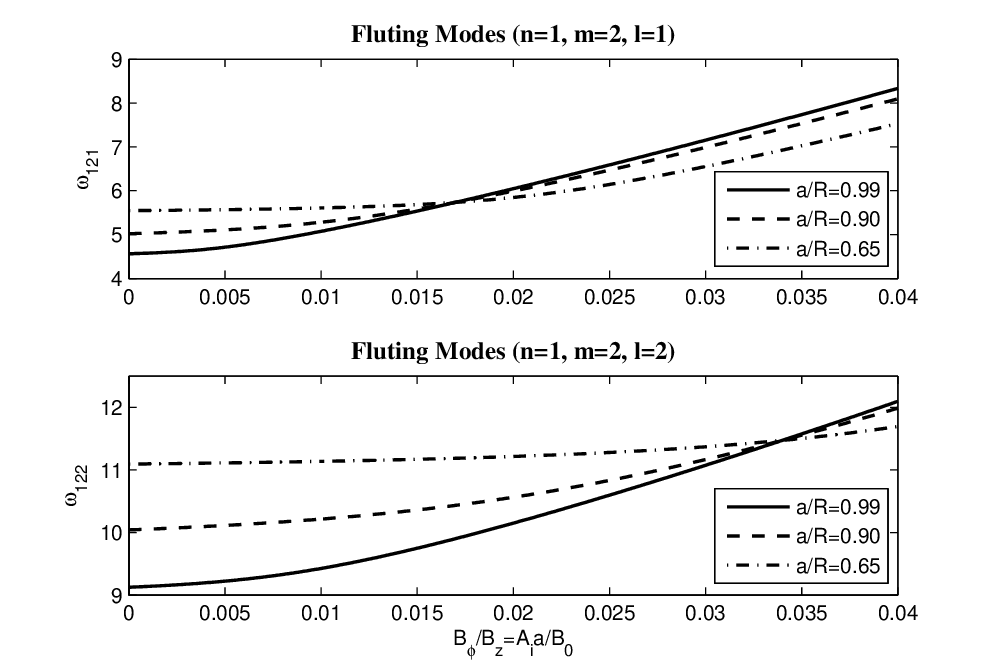}
      \vspace{6cm}
      \caption[]{Frequencies of the fundamental and its first overtone fluting $(m=2)$ modes with radial mode number $n=1$
      versus the internal twist parameter, $B_{\phi}/B_z=\frac{A_{\rm i}a}{B_0}$, for different
relative core widths $a/R=$ 0.65 (dash-dotted), 0.9
(dashed) and 0.99 (solid). The twist in the annulus region is
absent. Auxiliary parameters as in Fig. \ref{w1l1-w112}.}
         \label{w1l1-w112-m2}
   \end{figure}

As we mentioned before, the internal twist does not affect the kink
$(m=1)$ modes. Hence, we extend our investigation to the fluting
$(m=2)$ modes and study the effect of internal twist on their
frequencies by the numerical solution of the dispersion relation,
Eq. (\ref{dp}), in the absence of twist in the annulus region.
Figure \ref{w1l1-w112-m2} shows the frequencies of the fundamental
and first overtone $l=1,2$ fluting $(m = 2)$ modes with radial mode
number $n=1$ versus the internal twist parameter,
$B_{\phi}/B_z:=\frac{A_{\rm i}a}{B_0}$, when $A_0=0$ and for
different relative core widths $a/R=(0.65, 0.9, 0.99)$. Figure
\ref{w1l1-w112-m2} clears that: i) for a given $a/R$, the
frequencies increase when the internal twist parameter increases.
ii) For a given $a/R$, when the longitudinal mode number, $l$,
increases, the frequencies increase. iii) When $a/R$ goes to unity
then the frequencies obey Eq. (\ref{omega}).

\begin{figure}
\includegraphics{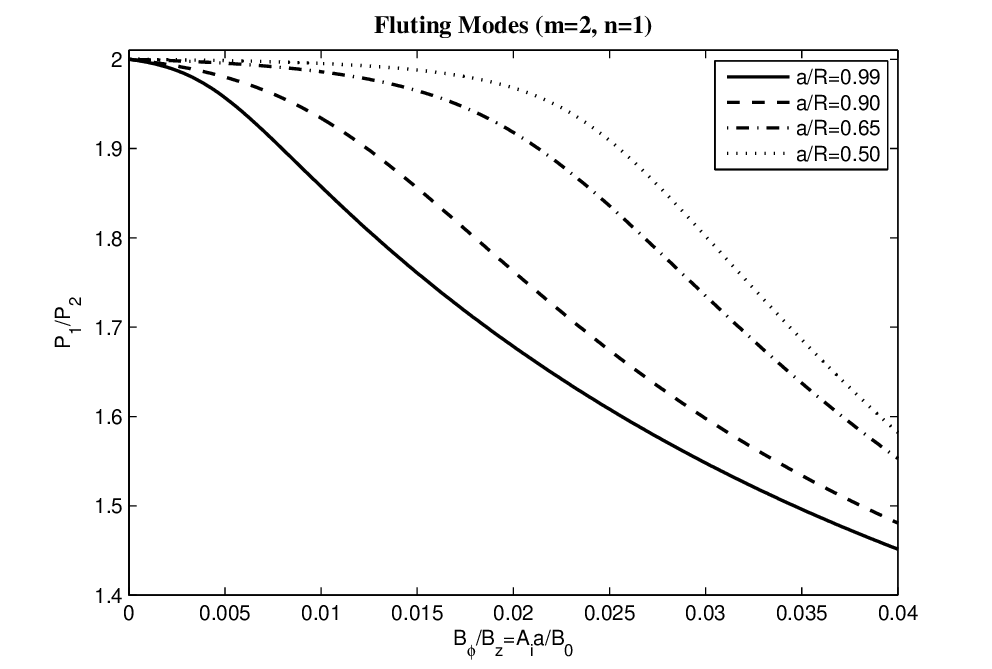}
      \vspace{6.4cm}
\caption[]{The period ratio $P_1/P_2$ of the fundamental and its
first overtone fluting $(m=2)$ modes with radial mode number $n=1$
versus the internal twist parameter, $B_{\phi}/B_z=\frac{A_{\rm
i}a}{B_0}$, for different relative core widths $a/R=$ 0.5 (dotted),
0.65 (dash-dotted), 0.9 (dashed) and 0.99 (solid). The twist in the
annulus region is absent. Auxiliary parameters as in Fig.
\ref{w1l1-w112}.}
         \label{n1-p1p2-m2}
   \end{figure}

Figure \ref{n1-p1p2-m2} shows the period ratio $P_1/P_2$ of the
fundamental and first overtone, $l=1,2$ modes of the fluting $(m=2)$
waves with $n=1$ versus the internal twist parameter when $A_0=0$.
Figure \ref{n1-p1p2-m2} presents that: i) the period ratio $P_1/P_2$
with increasing the internal twist parameter, decreases from 2 (for
untwisted loop) and approaches below 1.6 for $a/R=0.5$, for
instance. ii) For a given $B_{\phi}/B_z$, the period ratio $P_1/P_2$
decreases when the relative core width increases.

%-----------------------------------------------------------------------------------------------
\section{Conclusions}\label{Con}
Oscillations of nonaxisymmetric MHD waves in coronal loops in the
presence of the twisted magnetic field is studied. To do this, a
coronal loop is considered as a straight cylindrical compressible
zero-beta thin flux tube with magnetic twist in the internal and the
annulus and straight magnetic field in the external region. Using
the perturbation method given by Ruderman (2007), the dispersion
relation is obtained and solved both analytically and numerically
for obtaining the frequencies of the nonaxisymmetric modes. Our
dispersion relation confirms the results of other people for the
different cases. For instance: i) it shows that in the absence of
annulus region, the twist does not affect the kink ($m=1$) modes
which is same as the result obtained by Goossens, Hollweg $\&$
Sakurai (1992) and Ruderman (2007). ii) In the absence of internal
twist, the dispersion relation reduces to the same result derived by
Carter $\&$ Erd\'{e}lyi (2008) in the TT approximation. iii) In the
absence of twist in the annulus region, the dispersion relation
yields the same result obtained by Carter \& Erd\'{e}lyi (2007) for
the kink ($m=1$) modes in the TT approximation. Furthermore, the
effect of the internal twist on the fluting ($m=2$) modes is
investigated. Our numerical results show that

i) for a given relative core width, frequencies of the fundamental
and first overtone $l=1,2$ kink $(m=1)$ modes with radial mode
numbers $n=1,2$ increase when the twist parameter of the annulus
increases. The same behavior holds for the frequencies of the
fluting $(m=2)$ modes with $n=1$ when the internal twist parameter
increases;

ii) when the relative core width, $a/R$, goes to unity then the kink
$(m=1)$ modes with $n=1$ become independent of the twist and in the
case of $a/R=1$ the second mode labeled by $n=2$ is removed from the
system;

iii) the period ratio $P_1/P_2$ for the kink $(m=1)$ modes with
$n=1,2$ is lower than 2 (for untwisted loop) in the presence of the
twisted magnetic annulus. The results of $P_1/P_2$ for the kink
$(m=1)$ modes with $n=1$ are in accordance with some observations of
the TRACE. The period ratio for the fluting $(m=2)$ modes with $n=1$
decreases from 2 with increasing the internal twist parameter.

%--------------------------------------------------------------------------------------------------------------------
\section*{Acknowledgements}
The authors thank the anonymous referee for a number of
valuable suggestions. The work of K. Karami has been supported
financially by Research Institute for Astronomy and Astrophysics of
Maragha (RIAAM) under research project No. 1/1551.

%--------------------------------------------------------------------------------------------------------------------

%-----------------------------------------------------------------------------------------------
\clearpage
%-----------------------------------------------------------------------------------------------
%\clearpage
%-----------------------------------------------------------------------------------------------
\clearpage
%-----------------------------------------------------------------------------------------------
%\clearpage
%-----------------------------------------------------------------------------------------------
\clearpage
%-----------------------------------------------------------------------------------------------
%\clearpage
%-----------------------------------------------------------------------------------------------
\clearpage
%-----------------------------------------------------------------------------------------------
%\clearpage
%-----------------------------------------------------------------------------------------------
\clearpage
%-----------------------------------------------------------------------------------------------
%\clearpage
%-----------------------------------------------------------------------------------------------
\clearpage
\begin{table}
\centering\caption[]{Coronal seismology using the period ratio
$P_1/P_2$ of the kink modes $(m=1,n=1)$: the twist parameter of the
annulus $B_{\phi}/B_z=\frac{A_0a}{B_0}$ and the relative core width
$a/R$.}
\begin{tabular}{lccc}\hline\noalign{\smallskip}
 & $P_1/P_2$ &$B_{\phi}/B_z=A_0a/B_0$ &$a/R$
\\\hline\noalign{\smallskip}
Van Doorsselaere et al. (2007) & $1.795\pm 0.051$ & $0.0097$,
$0.0204$ &
$0.35$\\\\
%Van Doorsselaere et al. (2007) & $1.82\pm 0.08$ & $0.0107$, $0.0153$ & $0.50$\\
%McEwan et al. (2008) &  &  & \\\\
Van Doorsselaere et al. (2009) & $1.980\pm0.002$ & $0.0014$,
$0.0219$ &
$0.92$\\\\
Ballai et al. (2011) & $1.82\pm0.02$ & $0.0079$, $0.0231$ & $0.35$\\
\hline
\end{tabular}
\label{table}\\
\end{table}
%-----------------------------------------------------------------------------------------------
\end{document}